\documentclass[lettersize,journal]{IEEEtran}
\usepackage{amsmath,amsfonts}
\usepackage{mathtools}
\usepackage{algorithmic}
\usepackage{algorithm}
\usepackage{array}
\usepackage[caption=false,font=normalsize,labelfont=sf,textfont=sf]{subfig}
\usepackage{textcomp}
\usepackage{stfloats}
\usepackage{url}
\usepackage{verbatim}
\usepackage{graphicx}
\usepackage{cite}
\usepackage{hyperref}
\begin{document}

\title{HermEIS: A Parallel Multichannel Approach to Rapid Spectral Characterization of Neural MEAs}

\author{Akwasi Akwaboah,~\IEEEmembership{Student Member,~IEEE,}\quad
Ralph Etienne-Cummings,~\IEEEmembership{Fellow,~IEEE,}

\thanks{A. Akwaboah and R. Etienne-Cummings are with the Department of Electrical and Computer Engineering, Johns Hopkins University, Baltimore MD, 21218, USA. email: \{aakwabo1, retienne\}@jhu.edu. This manuscript has been submitted to the IEEE EMB Conference 2024}
}



\maketitle

\begin{abstract}
The promise of increasing channel counts in high density ($> 10^4$) neural Microelectrode Arrays (MEAs) for high resolution recording comes with the curse of developing faster characterization strategies for concurrent acquisition of multichannel electrode integrities over a wide frequency spectrum. To circumvent the latency associated with the current multiplexed technique for impedance acquisition, it is common practice to resort to the single frequency impedance measurement (i.e. $Z_{1 \text{kHz}}$). This, however, does not offer sufficient spectral impedance information crucial for determining the capacity of electrodes at withstanding slow and fast-changing stimulus and recordings. In this work, we present \textit{HermEIS}, a novel approach that leverages single cycle in-phase and quadrature signal integrations for reducing the massive data throughput characteristic of such high density acquisition systems. As an initial proof-of-concept, we demonstrate over $6$ decades of impedance bandwidth ($5\times10^{-2} - 5\times10^{4}\text{ Hz}$) in a parallel $4$-channel potentiostatic setup composed of a custom PCB with off-the-shelf electronics working in tandem with an FPGA.
\end{abstract}



\section{Introduction}
Ever since Santiago Ramón y Cajal's \textit{neuron doctrine}, which posited the nervous system as a conglomeration of discrete neurons, scientists and engineers continue to push the resolution of neural microelectrode arrays (MEAs) beyond single neuron dimensions (i.e. a few microns). Similar to CMOS imagers, higher electrode resolution allows us to perceive finer electrophysiology. Although there may exist other noninvasive interfacing modalities, direct electrochemical interfacing offers better temporal contrast to catch the rapidly propagating neural action potential. Finer electrode diameter and pitch are often accompanied with high MEA channel counts, which are advantageous in capturing the high-dimensional neural/ cortical processing of sensory information. Unfortunately, an increased channel density in a recording setup means that the digital processor has to drink from a fire hose of biosignals. Faster communication and processing are thus needed to enable rapid closed-loop control. The situation is much dire in spectral electrode characterization methods, such as Electrochemical Impedance Spectroscopy (EIS), needed to guarantee electrode integrity. Neural MEAs interfacing with tissue form a Helmhotz double layer, i.e. a capacitive double layer formed by charge redistribution induced by the electrode-electrolyte complex. Ideally, charge must travel from electrode to tissue (stimulation) and vice versa (recording) in either a capacitive non-faradiac manner or a reversible faradiac (redox) manner. This process is, however, limited by corrosion and glial cells formed over time. The capacitive nature of electrode/interface impedance requires that an investigator not only resolve for space and time but also for frequency. State-of-the-art MEAs have $>10^4$ channel counts, e.g., Neuropixels 2.0 ($10,240$) \cite{steinmetz2021neuropixels}, MaxOne/ MaxTwo (Maxwell Biosystems) ($26,400$) \cite{muller2015high} and Dragas \textit{et al.}\cite{dragas2017vitro} ($59,760$). It is common practice to circumvent the prolonged impedance acquisition associated with such high-density MEAs with mere measurements at $1$ kHz, $Z_{1\text{kHz}}$. While this may offer coarse interface information, the nuances that wide spectral impedance offer is crucial to designing durable electrode. 

To obtain such rich spectral information, one may choose to apply several cycles per test frequency sufficient to resolve the magnitude and phase information by Fourier transformation. This is time-consuming. More so, present characterization systems (e.g.  Gamry Instruments) handle multichannel characterization by multiplexing a cumbersome rack of potentiostats and thereby extends latency in data acquisition. In this study, we present \textit{HermEIS}, a rapid EIS chatacterization approach (see Fig. \ref{fig:setup}A) based on in-phase and quadrature integrations where a single frequency cycle suffices. We adopted a parallel multichannel acquisition in a Field Programmable Gate Array (FPGA), limited only by universal serial bus (USB) transmission to the host PC. As a proof-of-concept, our demonstration involves an $4$-channel Analog Front-End (AFE) implemented in a custom PCB working in tandem with the FPGA. The concepts presented here are however scalable to large-scale neural MEAs especially with the ever increasing speeds ($>10$ Gbit/s) in PC-peripheral communication (e.g. USB 3x--4.x, PCIe 4.x--5x).

In a potentiostatic setup (i.e. voltage in, current out), EIS involves the measurement of impedance magnitude and phase responses via the application of sinusoidal potentials with frequencies within $10^{-3} - 10^5$ Hz and root-mean-square (rms) magnitudes typically below $50$ mV to ensure current-voltage linearity \cite{cogan2008neural, merrill2005electrical}; as such interfacial impedance are nonlinear in a large signal sense.  The broad frequency spectrum mentioned above allows for a proper assessment of the dynamic range of electrode response. This is particularly important in guaranteeing the electrodes can capture the slowly changing subthreshold neural dynamics as well as the sharp neural depolarizations. The various frequency regimes of an EIS measurement indicate the quality of the electrode-electrolyte interface. While impedance response can be fitted a number of equivalent circuits of the electrolyte-electrolyte interface, the Randles circuit is perhaps among the simplest yet plausible \cite{randles1947kinetics, park2003peer}. It comprises the double layer capacitance, $C_{dl}$ in parallel with Faradiac impedance, $Z_F$ (composed of the redox kinetics resistance in series with the Warburg impedance, which accounts for the mass transport of the redox species), all in series with the uncompensated solution resistance, $R_U$. At both extremes, the interface impedance is dominated by $R_U$ and $\Re\{Z_W\}$ at low frequency, and dominated by the reactances of $C_{dl}$ and $Z_F$ at high frequencies. Thus, an impedance response in the form of Nyquist or Bode plots offer a finer insight on the state of electrodes when placed \textit{in vivo} or \textit{in vitro}. This work shares similarities with lock-in amplifier approach, in the sense that only a single period is sufficient. The difference, however, is in the sole-dependence on quarter cycle integrations here without needing any form of modulation, while lock-in amplifiers and other classical single cycle fourier techniques require a multiplication with a reference square wave of unity amplitude and cosine and sine modulators respectively \cite{orazem2008electrochemical}.

\section{Theory and Approach}
In a 3-electrode potentiostat setup, that is, the reference electrode (REF), counter electrode (CE), and working electrode (WE), the impedance $Z(\omega_i)$ is a function of the sinusoidal voltage, $V (\omega_i,t)$ applied on WE via REF and the WE current, $I(\omega_i, t)$ where $\omega_i$ is a user-defined frequency. Assuming linear time-invariance holds, the impedance response will either modulate amplitude $V_1$ or phase $\phi$ only. As such consider that $V(\omega_i,t)$ and $R_{\text{out}}I(\omega_i, t)$ are of the form,
\begin{equation}
    \label{eq:X_sig}
    V(\omega_i,t) = V_0 + V_1\sin{(\omega_i t + \phi)}
\end{equation}
where $V_0$ is an applied DC offset to set the operating point on the nonlinear IV curve and $R_\text{out}$ is a suitably chosen output resistance to convert $I(\omega_i,t)$ into a voltage to match voltage-mode A/D conversion used for the reference voltage, $V^{\text{(ref)}}$. The goal is to decipher phase and magnitude change information and there are a number of ways to obtained that. The proposed method adopts the I/Q approach not by the typical correlation of the WE output with a separately generated complimentary sinusoid pair, but rather via appropriate quarter cycle signal integrations shown in eqs.\ref{eq:I_cont},\ref{eq:Q_cont} over a period, $T=2\pi/\omega_i$.
\begin{equation}
    \label{eq:I_cont}
    \mathcal{I}(\omega_i) = \int_0^{T/2} {V(\omega_i, t)}dt - \frac{1}{2}\int_0^{T} {V(\omega_i,t)}dt \hspace{0.5em} =\hspace{0.5em} \frac{2V_1}{\omega_i}\cos{\phi}
\end{equation}
\begin{equation}
\label{eq:Q_cont}
    \mathcal{Q}(\omega_i) = \int_{T/4}^{3T/4} {V(\omega_i, t)}dt - \frac{1}{2}\int_0^{T} {V(\omega_i,t)}dt = -\frac{2V_1}{\omega_i}\sin{\phi}
\end{equation}
Let $\mathcal{X}(\omega_i)$ be an intermediary signal such that;
\begin{equation}
    \label{eq:X_int}
    \mathcal{X}(\omega_i) = \mathcal{I}(\omega_i) - j\mathcal{Q}(\omega_i)
\end{equation}
Note that $\mathcal{X}^{(\text{ref})}(\omega_i)$ takes the form $\mathcal{X}(\omega_i)$, while $\mathcal{X}^{(\text{ch}_j)}(\omega_i)$ takes the form $-\mathcal{X}(\omega_i)$ to account for negative gain in the WE current readout amplification stage.
The impedance response of a WE, $\text{ch}_j$, in an MEA of $j$ channels can then be;
\begin{equation}
    \label{eq:Zmag}
     \vert Z^{(\text{ch}j)}(\omega_i)\vert = R_\text{out}\cdot\left \vert \frac{ \mathcal{X}^{(\text{ref})}(\omega_i) }{\mathcal{X}^{(\text{ch}j)}(\omega_i)}\right\vert
\end{equation}
\begin{equation}
    \label{eq:Zang}
     \angle Z^{(\text{ch}j)}(\omega_i) =  \angle \mathcal{X}^{(\text{ref})}(\omega_i) - \angle \mathcal{X}^{(\text{ch}j)}(\omega_i)
\end{equation}

A discrete-time (DT) signal perspective on the above-mentioned is necessary for a digital implementation. In this case, the user-defined frequency $\Omega_i$ is posed as a function of the maximum ADC sampling frequency, $f_s$ in the form $\Omega_i = 2\pi f_i/f_s$, By sampling theorem, $\Omega_i\in (0,\pi)$, thus $\vert f_i/f_s\vert < 0.5$ and $f_{max}=f_s/2$. The DT version of eq.\ref{eq:X_sig} shown in eq.\ref{eq:X_sig_dis} is generated by direct digital synthesis (DDS) based on a clock frequency $f_{dds,clk}$. 
\begin{equation}
    \label{eq:X_sig_dis}
    V[\Omega_i,n] = V_0 + V_1\sin{(\Omega_i n + \phi)}
\end{equation}
To ensure periodicity and minimize approximation errors in the quarter cycle integrations, an adaptive sampling frequency, $f_s^{'}$ as a function of the $f_i$ is adopted and is defined over $f_i \in [f_{max}^{-1}, f_{max}]$. To get the most out of a given $f_s$, we adopted an adaptive sampling:
\begin{equation}
    \label{eq:fs_p1}
    f_s^{'}(f_i) = 
    \begin{cases}
        \left\lfloor \frac{f_s}{f_i} \right\rfloor \cdot f_i & \text{if } \left\lfloor \frac{f_s}{f_i} \right\rfloor \bmod {4}=0\\
        \left\lfloor \frac{f_s}{4f_i} \right\rfloor\cdot 4f_i & \text{elsewhere }  \\
    \end{cases}
\end{equation}
{\centerline {$\forall f_i\in (0, \lfloor \frac{f_s}{4} \rfloor]$}}
where $\lfloor\cdot\rfloor$ is flooring operation. Another round of approximation (shown in eq.\ref{eq:k0}) is performed to ensure that $f_s^{'}$ is an integer multiple of $f_{clk}$. $k$ is the clock divider for achieving this. 
\begin{equation}
    \label{eq:k0}
    k = \left\lfloor\frac{f_{clk}}{f_s^{'}}\right\rceil
\end{equation}
This rounding operation ($\lfloor\cdot\rceil$) introduces a marginal frequency, $f_{\epsilon}$ shift that is kept negligible by ensuring that $f_{clk}>>f_s^{'}$. The effective sampling frequency, $\hat{f_s}$ is thus given by:
\begin{equation}
    \hat{f}_s = kf_{clk} \pm \overbrace{\frac{f_s^{'}}{f_{clk}}}^{f_\epsilon}
\end{equation}

At this point, we adopt $\hat{\Omega}_i$ instead of $\Omega_i$, where $\hat{\Omega}_i = 2\pi f_i/\hat{f_s}$. DT equivalents of eqs.\ref{eq:I_cont},\ref{eq:Q_cont} with sampling period, $\Delta t_s=\hat{f}_s^{-1}$ and discrete time, $n$ are shown in eqs.\ref{eq:I_dis},\ref{eq:Q_dis} respectively. 

\begin{equation}
    \label{eq:I_dis}
    \mathcal{I}[\hat{\Omega}_i] = \left(\sum_0^{T/2} {V[\hat{\Omega}, n]} - \frac{1}{2}\sum_0^{T} {V[\hat{\Omega}_i,n]}\right)\cdot \Delta t_s
\end{equation}
\begin{equation}
    \label{eq:Q_dis}
    \mathcal{Q}[\hat{\Omega}_i] = \left(\sum_{T/4}^{3T/4} {V[\hat{\Omega}_i, n]} - \frac{1}{2}\sum_0^{T} {V[\hat{\Omega}_i,n]}\right)\cdot \Delta t_s
\end{equation}
\section{Implementation}
We implemented the above approach in an Opal Kelly (OK) XEM6310-LX150 FPGA working in tandem with an $4+1$-channel (i.e., 4 WEs and a REF) AFE implemented in a custom PCB with off-the-shelf electronics. The system is accompanied by a software controller in a form of python code with the \href{https://opalkelly.com/products/frontpanel/}{OK Frontpanel API} support. This allows for user-defined hardware initialization along with input/output (IO) data stream from the FPGA. The AFE was, however, designed for up to 8 channels, but only 5 were used based on maximum slice utilization in the FPGA (run at $f_{clk}=50$MHz to meet timing).

\begin{figure*}
    \centering
    \includegraphics[width=\textwidth]{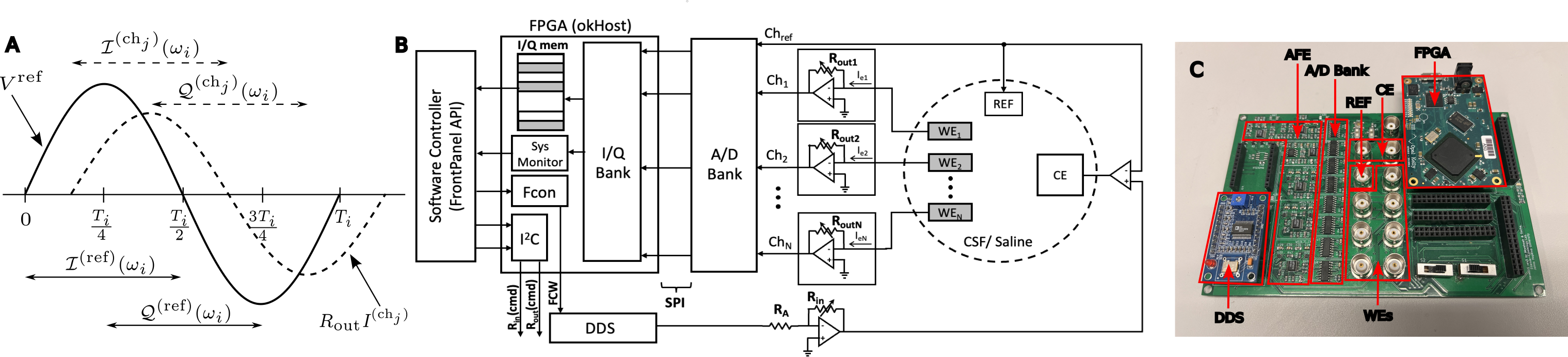}
    \caption{\textbf{A}: An intuitive depiction of the integrative intervals yielding I/Q pairs for the reference and WE channels, both subsequently used to obtained WE impedances. \textbf{B}: An overview of the hardware setup that combines RTL logic in an FPGA and analog front circuitry intended for multichannel \textit{in vivo}/ \textit{in vitro} measurements in a custom PCB shown in \textbf{C}}
    \label{fig:setup}
\end{figure*}

\subsection{Analog Front-End (AFE)}
In our multichannel setup shown in Fig. \ref{fig:setup}B, all WEs share a REF and a CE. We implemented the $8$-channel potentiostat using $2+8$ opAmp configuration, i.e., $2$ opAmps are designated for the DC offset addition and setting $V^{\text{ref}}$ respectively, and $8$ opAmps (one per channel) for conveying the output currents. $3$ LTC6082 (Analog Devices) Quad opAmp IC sufficed for this. We used the $8$ 10-bit SAR MCP3008 (Microchip Technology) ADCs that allows a configurable sampling rate of up to $200$ ksps sufficient for up to $50$ kHz impedance characterization. Each ADC comes with 8 single-ended channels. For each ADC, we connected channel-$0$ of $\text{ADC}_j$ to $\text{ch}_j$ WE and populated the other ADC channels with the remaining WE channels in a ring manner. As such, parallel acquisition from all WEs is possible by broadcasting a single channel address to all ADCs. Alternatively, a pseudo-parallel acquisition of all 8 channels is possible with one ADC by multiplexing the channel addresses. However, this comes with a reduced sampling rate at $1/8$ factor. More so, temporal shifts in sampling for this case must be accounted for. A combination of the parallel and pseudo-parallel modes stand to offer streaming from much higher channel counts. However, a commensurately high sampling rate is required. Here, we focus on the parallel mode and reserve the other mode for future study. Single-ended ADC inputs were used. This required elevating the measured potentials well above ground to allow the ADC to catch the full signal swing. An offset of $V_{dd}/2 \approx 1.65$ V was applied here. $V_0$ of the $V^{\text{ref(i)}}$ is set to $0$ through an opamp-based inverting adder with an appropriate offset. The final reference voltage $V^{\text{ref(o)}}$ can be amplitude modulated via $R_{\text{in}}$ in the same amplification stage defined by $V^{\text{ref(o)}}=-(R_{\text{in}}/R_{\text{A}})V^{\text{ref(i)}}$ where $R_A=100k\Omega$.
We used the AD9850 (Analog Devices) DDS module for signal generation. This module conveniently comes with a $42$ MHz low-pass, $5$-pole elliptical
filter at its DAC output and it is capable of up to $40$ MHz output. Alternatively, a numerically controlled oscillator can be synthesized in an FPGA with a sine Look-Up Table (LUT) along with an appropriate output digital-to-analog converter (DAC) and low-pass filter (LPF) if tighter system integration is desired.

\subsection{FPGA Implementation and Software Controller}
For every $f_i$ specified via a $M$-bit frequency control word (FCW) $m$ based on a DDS clock, $f_{dds,clk}=100$ MHz (given in Eq.\ref{eq:fcw}), the FPGA computes the appropriate $\hat{f}_s$ and retrieves a cycle-long stream of $10$-bit sampled data via the serial peripheral interface (SPI) protocol at $k$ count intervals.
\begin{equation}
    \label{eq:fcw}
    m = \frac{2^Mf_i}{f_{dds,clk}}
\end{equation}
To minimize duplicate integrations, we express eq.\ref{eq:I_dis},\ref{eq:Q_dis} in terms of quarter cycle integrations $S_j$ as shown in eq. \ref{eq:I_fp},\ref{eq:Q_fp}.
\begin{equation}
    S_j = -\frac{r_jX_{\alpha,j}}{4m} + \sum_{n = \left\lfloor \frac{jT}{4} \right\rfloor}^{\left\lfloor \frac{(j+1)T}{4} \right\rfloor} {X[n]} -\frac{(4m - r_{j+1})X_{\alpha,j+1}}{4m}
\end{equation}
where $r_j = j\frac{2^M\hat{f}_s}{f_{dds_clk}} \mod 4m$, $X$ is the $10$-bit ADC data, $X_{\alpha,j}=\left\lfloor \frac{(j+1)T}{4} \right\rfloor$ and $j \in \{0,1,2,3\}$

\begin{equation}
    \label{eq:I_fp}
    \mathcal{I}[\hat{\Omega}] = ( S_0 + S_1 - (S_2 + S_3) )/{2\hat{f}_s}
\end{equation}
\begin{equation}
    \label{eq:Q_fp}
    \mathcal{Q}[\hat{\Omega}] = ( S_1 + S_2 - (S_0 + S_3) )/{2\hat{f}_s}
\end{equation}
We compute $(\mathcal{I}, \mathcal{Q})\cdot\hat{f}_s$ equivalent in the FPGA and defer the determination of magnitude and phase responses to the software controller which queries for this I/Q pair. 
We use $7$-bit digital rheostats (MCP40D17, Microchip Technology) to for setting $R_{\text{in}}, R_{\text{out},j}$ defined by;
\begin{equation}
    \label{eq:digi_rheo}
    R_{\{\text{in,out}\}} = R_{min} + ({R_{max}N_{\text{(in,out)}}})/{127}
\end{equation}
where is $N\in[0,127]$ is the 7-bit a user-defined command word set in the software controller and delivered by the FPGA via $\text{I}^2\text{C}$, and $R_{max}=50k\Omega$ and $R_{min} = 0.002R_{max}$ (typical value reported in datasheet) being the maximum and minimum resistances respectively.
This ensured the flexibility of finding appropriate resistor values that promote high SNR without clipping signals. We used a 32-bit signed fixed-point precision to achieve a 6-decade impedance spectra ($5\times10^{-2}$$ - 5\times10^{4}$ Hz) computation, limited at the lower bound by DDS resolution ($f_{dds,clk}/2^{31}$) and the upper bound by sampling limit of $f_s/4$. A $32\times10$-bit buffer is instantiated in the FPGA for temporarily staging the 5 I/Q pairs. With the help of the software controller, in the form of a jupyter notebook, and a programmed FPGA; impedance acquisition process proceeds as follows. The user specifies a list of frequency points to be computed and loops through these points while grabbing I/Q pair data at every iteration. An iteration ends once the system monitor in the FPGA signals the completing of a signal cycle. This signaling occurs at the end of second cycle of I/Q generation. Two cycles instead of one allows an overwrite of any noise injection that occurs in the first cycle due to the switch in frequency, albeit scaling the acquisition time by a factor of 2. Regardless, this may still be significantly time relative to other fourier cycles requiring several cycles per frequency to resolve impedance.

\begin{figure*}
    \centering
    \includegraphics[width=0.9\textwidth]{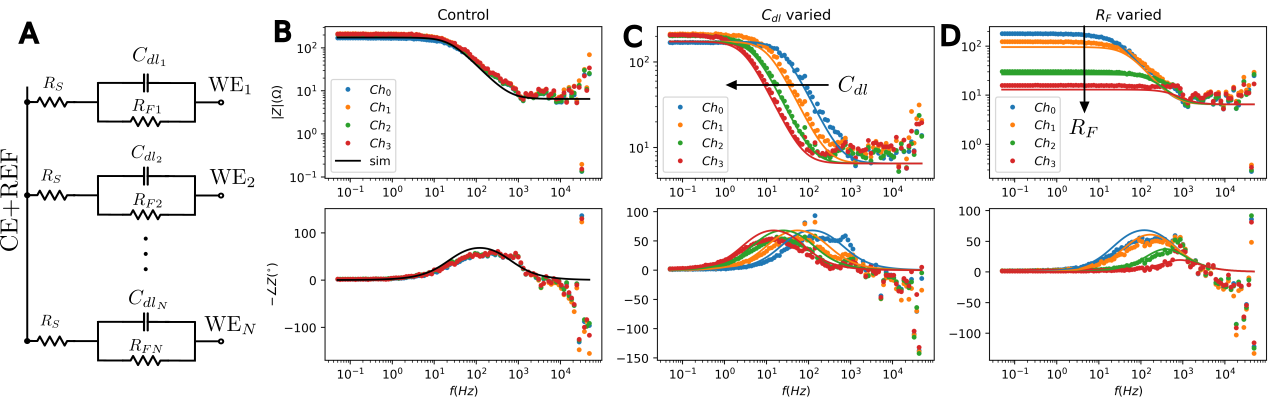}
    \caption{Bode plots from parallel $4$-channel impedance acquisition in 3 experimental protocols along with simulated ground truth in solild lines -- \textbf{B}: Control ($R_S = 3.9k\Omega,\ R_F = 100k\Omega,\ C_{dl}=0.068\mu F$), \textbf{C}: Varying $C_{dl}$, $R_F=100k\Omega$, $C_{dl} = \{0.068, 0.15, 0.33, 0.56\} \mu F$ for $\text{Ch}_{1,\cdots,4}$ respectively, \textbf{D}: Varying $R_{F}$, $C_{dl} = 0.068\mu F$, $R_F = \{100, 53.6, 12, 3.9\}k\Omega$ for $\text{Ch}_{1,\cdots,4}$ respectively. $\alpha=1/600$ calibration factor applied.}
    \label{fig:results}
\end{figure*}

\section{Results and Discussion}
For test purposes, we modeled each electrode interface with a simplified Randles equivalent circuit composed of a series Resistance, $R_S,$ connected to a parallel combination of resistance $R_F$ (intended to model the Faradaic process) and capacitance $C_{dl}$ (intended to model the interfacial capactive double layer) in Fig \ref{fig:results}. This exhibits high-pass filtering with a cut-off frequency, $\omega_C=1/(R_F C_{dl})$ and an $R_S$ offset as shown in eq. \ref{eq:imp_mod}.
\begin{equation}
    \label{eq:imp_mod}
    Z_o(\omega) = R_S + \frac{R_F}{1+j\omega R_F C_{dl}}
\end{equation}
A two-electrode configuration with CE and REF tied at one end of the impedance terminal and the other to a WE on a breadboard with jumper wires. Two test setups were explored -- varying $R_F$ and $C_{dl}$ respectively. Arbitrary component values were chosen here. However, the intent was to depict how possible neural electrode-electrolyte interfacial impedance variations (arising from biotic and abiotic changes, e.g. corrosion and glial cell formation) can be concurrently captured over multiple electrode channels and as wide a spectrum as possible. Results for this can be seen in Fig \ref{fig:results}. In the control setup, the following RC values are used -- $R_s = 3.9k\Omega$, $R_F=100k\Omega$, and $C_{dl}=0.068\mu F$. In the $R_F$-varying experiment, $C_{dl} = 0.068\mu F$, $R_F = \{100, 53.6, 12, 3.9\}k\Omega$ for $\text{Ch}_{1,\cdots,4}$ respectively. While in the $C_{dl}$-varying experiment, $R_F=100k\Omega$, $C_{dl} = \{0.068, 0.15, 0.33, 0.56\} \mu F$ for $\text{Ch}_{1,\cdots,4}$ respectively. $N_{\text{in}}=10$ was used to set $V^{\text{ref(o)}}\approx 40 \text{ mVpp}$ below the typical $50\text{ mVpp}$ limit for ensuring small signal linearity. $N_{\text{out}}=100$ offers sufficient impedance SNR. The measured impedance is expressed as;
\begin{equation}
    \label{eq:Z_calib}
    Z(\omega) = \alpha Z_o(\omega)
\end{equation}
where $\alpha$ is a calibration factor accounting for component variations, possible non-linear IV relations arising from scaling $V^{\text{ref}}$ and $R_{out}$ scaling. We set $\alpha=1/600$ based on empirical observations for the above-mentioned protocols. A spectral scan with $100$ logspaced points over the ($5\times10^{-2} - 5\times10^{4}\text{ Hz}$) band took only $3$min $40$s. This comprises of $2\times$ periods for each frequency (with the largest period being $20$s), and the software controller overhead. Wider deviations are apparent at higher frequency due to approaching sampling limits.

\subsection{Scaling to More Channels}
The concept presented is scalable to high density neural MEAs bottlenecked mainly by the PC-FPGA communication speeds. The OK-XEM6310 comes with a USB 3.0 capable of up to $38.1$ MB/s transmission speeds for a block read length of 1024 bytes. Assuming a sufficiently large buffer for staging I/Q data in FPGA, an I/Q pair of $8$ bytes (i.e. $2\times32$ bits) precision, could support almost $5\times10^{6}$ electrode channels which is well beyond the $59,760$ state-of-the-art MEA channel density\cite{dragas2017vitro}. Of course, with the accompanying increased slice utilization which a newer and more capable FPGA may meet. More so, analog integration by log-domain filtering could potential reduce the high computational resource requirements.

\section{Future Work}
Moving forward, validation in \textit{in vitro} neural MEAs setups will be explored. This will be useful in capturing spectral shifts in impedance that arise due to aging and corrosive factors for various electrode materials. There exist opportunities to adopted nonlinear optimization methods to find ideal $R_{in}$ and $R_{out}$ values in the calibration process. Ultimately, an integrated circuit approach will allow for compactness and allow for low-resource I/Q pair generations as quarter cycle integrations could be performed with low-power analog subthreshold integrators. Code and supplementary made available at \href{https://github.com/Adakwaboah/HermEIS_IQ}{https://github.com/Adakwaboah/HermEIS\_IQ}.

\section*{Acknowledgments}
We would like to acknowledge Dr. James Weiland for invaluable insights while AA worked at his lab, and Jon Socha for help with PCB design.

{
\bibliographystyle{IEEEtran}
\bibliography{ms}
}

\end{document}